\documentclass[a4paper,fleqn]{cas-dc}

\usepackage[numbers]{natbib}

\def\tsc#1{\csdef{#1}{\textsc{\lowercase{#1}}\xspace}}
\tsc{WGM}
\tsc{QE}
\newcommand*\chem[1]{\ensuremath{\mathrm{#1}}}
\newcommand*{\pd}[3][]{\frac{\partial^{#1} #2}{\partial #3}}

\usepackage{subfig}
\usepackage{bigints}
\usepackage{bm}

\begin{document}
\let\WriteBookmarks\relax
\def\floatpagepagefraction{1}
\def\textpagefraction{.001}

\shorttitle{Hydrogen DNS}    

\shortauthors{Yao and Blanquart}  

\title [mode = title]{Isolating effects of large and small scale turbulence on thermodiffusively unstable premixed hydrogen flames}  



%

\author[1]{Matthew X. Yao}

\cormark[1]


\ead{mxyao@caltech.edu}



\affiliation[1]{organization={Department of Mechanical and Civil Engineering, California Institute of Technology},
            addressline={1200 E California Blvd}, 
            city={Pasadema},
            postcode={91125}, 
            state={California},
            country={USA}}

\author[1]{Guillaume Blanquart}






\cortext[1]{Corresponding author}



\begin{abstract}
Lean turbulent premixed hydrogen/air flames have substantially increased flame speeds, commonly attributed to differential diffusion effects. In this work, the effect of turbulence on lean hydrogen combustion is studied through Direct Numerical Simulation using detailed chemistry and detailed transport. Simulations are conducted at six Karlovitz numbers and three integral length scales. A general expression for the burning efficiency is proposed which depends on the conditional mean chemical source term and gradient of a progress variable.  At a fixed Karlovitz number, the normalized turbulent flame speed and area both increase linearly with the integral length scale ratio. The effect on the mean source term profile is minimal, indicating that the increase in flame speed can solely be attributed to the increase in flame area. At a fixed integral length scale, both the flame speed and area first increase with Karlovitz number before decreasing. At higher Karlovitz numbers, the diffusivity is enhanced due to penetration of turbulence into the reaction zone, significantly dampening differential diffusion effects.
\end{abstract}




\maketitle











\section*{Novelty and Significance}

Lean premixed hydrogen flames are subject to thermodiffusive instabilities, which can lead to system level instabilities such as flashback. A comprehensive study of the thermodiffusively unstable turbulent flames with detailed chemistry and detailed transport (including Soret diffusion) was conducted across a wide range of turbulent intensities and integral length scales. Varying these two parameters independently was necessary to isolate the effects of large and small scale turbulence. Using these results, we propose a general expression for the burning efficiency to explain the relationship between the turbulence intensity, flame speed, and flame area. This work is an important step in developing predictive models which can aid the design of practical combustion devices.

\section*{Author Contributions}
\textbf{M.X.Y.:} Conceptualization, methodology, software, validation, formal analysis, investigation, resources, writing -- original draft, visualization. \textbf{G.B.:} Conceptualization, methodology, writing -- review \& editing, supervision, funding acquisition

\section{Introduction}

Hydrogen combustion has emerged as a promising technology to mitigate the negative effects of fossil fuel combustion, such as greenhouse gas emissions and soot formation~\cite{verhelst2009hydrogen}. Hydrogen is typically burnt lean in order to reduce the flame temperature and hinder the formation of \chem{NO_x} pollutants. However, when burnt lean, hydrogen-air mixtures are susceptible to thermodiffusive instabilities, which arise due to the relatively low Lewis number (ratio of thermal to mass diffusivity) of hydrogen~\cite{sivashinsky1977diffusional}. Species with small molecular weight are also subject to Soret diffusion (mass diffusion driven by thermal gradients), which is known to enhance the instabilities~\cite{grcar2009soret}. The curvature of the flame front has a focusing effect on the differential diffusion, causing the fuel to become concentrated in some regions and dilute in others, resulting in local equivalent ratio fluctuations. The richer regions burn faster, and the leaner regions burn slower. 
In laminar flames, the instabilities manifest as characteristic cellular structures~\cite{bregeon1978near,mitani1980studies}, the onset of which are traditionally studied through linear stability analysis~\cite{matalon2003hydrodynamic,altantzis2012hydrodynamic,berger2022intrinsic}. When subject to turbulence, lean hydrogen flames exhibit highly enhanced flame speeds~\cite{kido2002influence} which introduce safety concerns such as flashback~\cite{verhelst2009hydrogen}. 

The special burning properties of turbulent lean hydrogen combustion have been investigated in a variety of Direct Numerical Simulation (DNS) studies using detailed chemistry. Ideally, simulations would be conducted in practical combustor configurations, for example the slot burner simulated by Berger \textit{et al.}~\cite{berger2022synergistic}. In these configurations, the flame is exposed to turbulence generated in the shear layer. Although physically realistic, these simulations are computationally expensive. As such, Berger \textit{et al.} have only simulated a single relatively low Karlovitz number, defined as the ratio of flame to turbulence timescales. Pushing to higher Karlovitz numbers requires a prohibitively expensive increase in computational cost.

To make the problem computationally tractable, the domain can be reduced to a doubly periodic inflow-outflow configuration to focus specifically on flame-turbulence interactions.  In these simulations, a forcing term is added to mimic the production of turbulence due to the missing large-scale mean shear~\cite{dhandapani2020isotropic}.  Blanquart and coworkers~\cite{savard2015broken,lapointe2015differential,bobbitt2016vorticityTrans} used this configuration extensively to investigate the two-way coupling between turbulence and combustion in neutrally stable hydrocarbon flames with detailed chemistry and transport.  Building on this work, Schlup and Blanquart~\cite{schlup2019reproducing} performed one of the earliest DNS of turbulent unstable hydrogen flames which incorporated detailed transport with Soret effects. However, they studied only a single Karlovitz number. 

The inflow-outflow configuration is a popular setup which has been used in a number of studies of lean turbulent hydrogen flames. Lee and coworkers~\cite{lee2021influence,lee2022dnsa} employed forcing through the whole domain to study the impact of leading points on flame speed enhancement in thermodiffusively unstable flames. However, they did not account for Soret effects. Song \textit{et al.}~\cite{song2021statistics} and Yuvraj \textit{et al.}~\cite{yuvraj2022local} conducted DNS and studied the statistics of global and local flame speeds. In particular, they showed a strong relationship between the turbulent flame speed and the integral length scale. Although their work included Soret diffusion, the simulations were conducted at an equivalence ratio of $\phi= 0.7$, rendering the flames neutrally stable. Thus, they were not able to capture the effects of thermodiffusive instability.

All of the listed studies thus far were conducted at low to moderate Karlovitz numbers. Aspden \textit{et al.}~\cite{aspden2011turbulence,aspden2019towards} conducted simulations at extreme Karlovitz numbers to study the transition to the distributed burning regime. Once again, their configuration consisted of a flame which propagated in a doubly-periodic box with isotropic turbulence maintained through long wavelength forcing. This work was furthered0 by Howarth \textit{et al.}~\cite{howarth2023thermodiffusively}, who conducted a detailed study on flame curvature and provided further support for leading point theory. Unfortunately, these studies did not include Soret diffusion.

The goal of the present work is to conduct a detailed and comprehensive study of lean turbulent hydrogen flames, incorporating Soret effects, and across a wide range of turbulence intensities. In particular, we aim to isolate local effects (e.g. flame structure) and global effects (e.g. turbulent flame area) and quantify their individual contributions to the turbulent flame speed enhancement.

The scaling of turbulent to laminar flame speed was first described by the phenomological or conceptual model of Damk{\"o}hler~\cite{damkohler1940einfluss}, who posited that locally, the flame front moves at the laminar flame speed. Thus, the following scaling was obtained:
\begin{equation}
    \frac{S_T}{S_L} = \frac{A_T}{A}.
\end{equation}
where $S_T$ is the turbulent flame speed, $S_L$ is the laminar flame speed, $A_T$ is the turbulent flame area, and $A$ is the domain cross-sectional area. As the turbulence wrinkles the flame, it necessarily increases the flame area, which can be considered a global quantity. Notably, this model is not able to account for the local effects of turbulence such as curvature or strain on the chemistry. The model was thus extended with the burning efficiency, $I_0$, to account for these effects~\cite{bray1990studies,candel1990flame}:
\begin{equation}
\label{eq:efficiency}
    \frac{S_T}{S_L} = I_0 \frac{A_T}{A}
\end{equation}
There is no consensus on a general expression for $I_0$. Based on scaling arguments from Savard and Blanquart~\cite{savard2015broken}, Lapointe and Blanquart~\cite{lapointe2016fuel} proposed the following scaling for the ratio of turbulent to laminar flame speed for hydrocarbon flames:
\begin{equation}
\label{eq:i0lapointe}
    \frac{S_T}{S_L} \approx  \frac{\left \langle\dot{\omega}_C/\vert \nabla C\vert \big \vert C_{peak} \right\rangle}{\dot{\omega}_{C,lam}/\vert \nabla C_{lam} \vert} \frac{A_T}{A}
\end{equation}
where $C$ is a progress variable, and $\dot{\omega}_C$ is the corresponding source term. This model proposes a method of calculating the burning efficiency, but relies on the assumption that the source term profiles locally scale as they do at $C_{peak}$. The applicability of this model has not yet been validated for hydrogen flames and is the subject of the present study.

In this paper, DNS are carried out using detailed chemistry and transport across a range of Karlovitz numbers and integral length scales to isolate effects of small-scale and large-scale turbulence. Sections~\ref{sec:problem} and~\ref{sec:forcing} describe the problem setup and numerical details. Section~\ref{sec:overview} provides an overview of the results. In Section~\ref{sec:glob_loc}, the global effects and the local effects are decoupled to deconstruct the various components of Eq.~\eqref{eq:efficiency} and develop a general expression for $I_0$. The local response of the flame is investigated in more detail in Section~\ref{sec:local}. The conclusions are drawn in Section~\ref{sec:conclusion}.

\section{Problem description}
\label{sec:problem}

In this section, the physical problem and numerical methodology are presented.

\subsection{Flow Configuration}

The inflow-outflow configuration~\cite{savard2015broken,savard2015structure,bobbitt2016vorticityIso,bobbitt2016vorticityTrans,lapointe2015differential,lapointe2016fuel,lee2021influence,lee2022dnsa,song2022diffusive} is commonly used to study turbulent premixed flames since it allows for the development of a statistically stationary, statistically planar, freely propagating flame without mean shear or strain. The computational domain is rectangular, with domain size $L_x \times L_y \times L_y$ with $N_x\times N_y \times N_z$ points in the $x,y,$ and $z$ directions respectively, where $x$ is the streamwise direction. The mean inlet velocity at $x=0$ is set to match the turbulent flame speed such that the flame is stationary within the domain. The boundary condition at $x=L_x$ is a convective outflow. In these simulations, $L_y = L_z = L$, and the domain is periodic in both the $y$ and $z$ directions. The grid resolution is the same in all three directions, that is, $\Delta x = \Delta y = \Delta z$. A detailed discussion of the turbulence generation is presented in the following sections. 

\subsection{Governing Equations}

In the present study, the variable density, low-Mach, reacting flow equations are solved. The conservation equations for mass, momentum, temperature, and species are written:

\begin{equation}
    \pd{\rho}{t} + \nabla \cdot \left(\rho \mathbf{u}\right) = 0
\end{equation}
\begin{equation}
    \pd{\rho \mathbf{u}}{t} + \nabla \cdot \left(\rho \mathbf{u} \otimes \mathbf{u} \right) = -\nabla p + \nabla \cdot \bm{\sigma} + \mathbf{f}
\end{equation}
\begin{align}
    \pd{\rho T}{t} + \nabla \cdot \left(\rho \mathbf{u} T\right) = \nabla \cdot \left(\rho \alpha \nabla T\right) + \dot{\omega}_T \nonumber \\  
    - \frac{1}{c_p} \sum_{i}^{n_s} c_{p,i} \mathbf{j}_i \cdot \nabla T + \frac{\rho \alpha}{c_p} \nabla c_p \cdot \nabla T
\end{align}
\begin{equation}
    \pd{\rho Y_i}{t} + \nabla \cdot \left(\rho \mathbf{u} Y_i\right) = -\nabla \cdot \mathbf{j}_i + \dot{\omega}_i
\end{equation}
In these equations, $t$ is the time, $\rho$ is the density, $\mathbf{u} = \{u,v,w \}$ is the velocity vector, $p$ is the hydrodynamic pressure, $T$ is the temperature, $c_p$ is the mixture heat capacity, and $\alpha = \lambda /(\rho c_p)$ is the mixture thermal diffusivity. The mixture-averaged thermal conductivity, $\lambda$, is calculated following Mathur \textit{et al.} ~\cite{mathur1967thermal} using species thermal conductivities obtained via Eucken's formula~\cite{eucken1913warmeleitvermogen}. The viscous stress tensor is given as:
\begin{equation}
    \bm{\sigma} = \mu \left(\nabla \mathbf{u} + \nabla \mathbf{u}^T\right) - \frac{2}{3} \mu \left(\nabla \cdot \mathbf{u}\right) \mathbf{I}
\end{equation}
where $\mu$ is the mixture dynamic viscosity and 
$\mathbf{I}$ is the identity tensor. The species viscosities are calculated following the standard gas kinetic relations~\cite{hirschfelder1954molecular}, and the corresponding mixture-averaged viscosity follows a modified form of Wilke's formula~\cite{lapointe2015differential,wilke1950viscosity}. The forcing term $\mathbf{f}$ in the momentum equation will be discussed in detail in the following section. For each species $i$, we have the mass fraction $Y_i$, the species heat capacity $c_{p,i}$, the species production rate $\dot{\omega}_i$ (units of [\chem{kg/m^3s}]), and the species mass diffusion flux $\mathbf{j}_i$.  The source term for the temperature equation is expressed as:
\begin{equation}
    \dot{\omega}_T = -\frac{1}{c_p} \sum_i^{n_s} h_i\left(T\right) \dot{\omega}_i
\end{equation}
where $h_i$ is the enthalpy of species $i$ at a given temperature.

In this work, the mixture-averaged diffusion model is employed. The species diffusion flux is then written: 
\begin{equation}
    \mathbf{j}_i = -\rho \frac{Y_i}{X_i} D_{i,m} \nabla X_i - \frac{1}{T} D_i^T \nabla T + \rho Y_i \mathbf{u}_c
\end{equation}
where for each species $i$, $D_{i,m}$ is the mixture-averaged diffusion coefficient, $D_i^T$ is the thermal diffusion coefficient, and $X_i$ is the mole fraction. To ensure mass conservation and zero net diffusion flux, a correction velocity is introduced~\cite{coffee1981transport,kee2005chemically}:
\begin{equation}
    \mathbf{u}_c = \mathbf{u}_c^D + \mathbf{u_c^T}
\end{equation}
which has contributions from the Fickian diffusion:
\begin{equation}
    \mathbf{u}_c^D = \frac{\nabla W}{W} \sum_i^{n_s} D_{i,m} Y_i + \sum_{i}^{n_s} D_{i,m} \nabla Y_i
\end{equation}
and the thermal diffusion:
\begin{equation}
    \mathbf{u}_c^T = \frac{1}{\rho} \frac{\nabla T}{T} \sum_{i}^{n_s}D_i^T
\end{equation}
where $W$ is the mixture molecular weight. The present simulations use the reduced model recently proposed by Schlup and Blanquart for the calculation of the thermal diffusion coefficients~\cite{schlup2018reduced,schlup2018validation}.

The equations are closed with the ideal gas equation of state:
\begin{equation}
    \rho = \frac{P_0 W}{RT}
\end{equation}
where $P_0$ is the background thermodynamic pressure, and $R$ is the universal gas constant.

\subsection{Numerical method}

The governing equations are solved using NGA~\cite{desjardins2008high}, which is a fully conservative, finite-difference, code of arbitrarily high spatial order. In the present work, the spatial and temporal discretization are both second-order accurate. The time integration is conducted via the semi-implicit iterative Crank-Nicolson method~\cite{pierce2001progress}. The WENO5~\cite{jiang1996efficient} scheme is used for scalar transport. To reduce the computational cost associated with integrating the stiff chemistry in time, the preconditioning strategy of Savard \textit{et al.}~\cite{savard2015computationally} is used. 

The grid resolution must be sufficient to fully resolve both the turbulence and chemistry. As such, $\Delta x$ is set to meet the more restrictive criteria between $k_{max} \eta_u > 1.5$~\cite{yeung1989lagrangian} and roughly 16 points per laminar flame thickness~\cite{schlup2018validation}. Here, $k_{max} = \pi/\Delta x$ is the maximum wavenumber which can be resolved on the grid. For hydrogen chemistry, Savard \textit{et al.}~\cite{savard2015computationally} showed that the maximum stable timestep size could be increased from $\Delta t_{max} = 5.2\times 10^{-8}$ using an explicit integration to $1.6\times 10^{-6}$ using the semi-implicit time integration. To ensure numerical stability, a slightly lower value was selected as the upper bound in the present simulations. As such, the timestep size is determined based on the more restrictive criteria between $\Delta t = 1\times 10^{-6}$ and the convective CFL, $\Delta t \leq 0.8 \Delta x/\max\left(\mathbf{u}\right)$. 

\subsection{Chemical model}

The fuel mixture consists of hydrogen/air at an equivalence ratio of $\phi = 0.4$. The chemistry is described by the 9 species 54 reaction mechanism proposed by Hong \textit{et al.}~\cite{hong2011improved} with a few updated rate constants~\cite{hong2013rate,lam2013shock}. The chemical model is provided in supplemental material. Some important quantities from the one-dimensional laminar unstretched freely propagating flame are summarized in Table~\ref{tab:laminar_parameters}. Consistent with typical studies of flamelets and flamelet modelling, the progress variable, $C$, is defined as the water mass fraction, $Y_\chem{H_2O}$. 

The flame thickness is defined as:
\begin{equation}
    l_F = \frac{T_b-T_u}{\max\left(\vert \nabla T \vert \right)}
\end{equation}
and $S_L$ is the laminar flame speed. The subscripts $u$ and $b$ represent the unburnt and burnt mixtures respectively. With thermal diffusion, the flame speed and peak source term are slightly increased, and the flame thickness is slightly decreased compared to the full transport. The thermal diffusion acts to push the \chem{H_2} and \chem{H} molecules into the preheat zone and away from the reaction zone, thereby slightly reducing the peak source term. For the unity Lewis case, the flame speed is about 2 times larger, and the peak source term is about 3.7 larger. The quantity $C_{peak}$ denotes the value of $C$ in the one-dimensional flame where the peak source term, $\dot{\omega}_{C,max}^{1D}$, is located.

Unless stated otherwise, the values used for normalization in the following sections are taken from the one-dimensional laminar flame with mixture average transport and thermal diffusion.

\begin{table}[tbh] 
\caption{Relevant parameters from one-dimensional laminar flames}
\centerline{\begin{tabular}{l| c c c}
Parameter         & Mix Avg TD   & Mix Avg no TD & Unity Le\\ \hline
$T_u$ $[K]$       & $298$        & $298$         & $298$ \\
$\phi$            & $0.4$        & $0.4$         & $0.4$ \\
$S_L$ $[m/s]$     & $0.206$      & $0.215$       & $0.410$ \\ 
$\delta$ $[mm]$   & $0.816$      & $0.812$       & $0.441$ \\
$l_F$ $[mm]$      & $0.683$      & $0.656$       & $0.375$ \\
$\nu_u$ $[m^2/s]$ & $1.62\times 10^{-5}$  & $1.62\times 10^{-5}$ & $1.62\times 10^{-5}$ \\
$C_{peak}$        & $0.0835$ & $0.0832$ & $0.0815$ \\ 
$C_{max}^{1D}$    & $0.1033$ & $0.1033$ & $0.1033$ \\ 
$\dot{\omega}_{C,max}^{1D}$ & $52.57$ & $54.84$ & $195.63$ \\ 
\end{tabular}}
\label{tab:laminar_parameters}
\end{table}

\subsection{Simulation parameters}

\begin{figure}[tbh]
    \centering
    \includegraphics[width=1.0\linewidth]
    {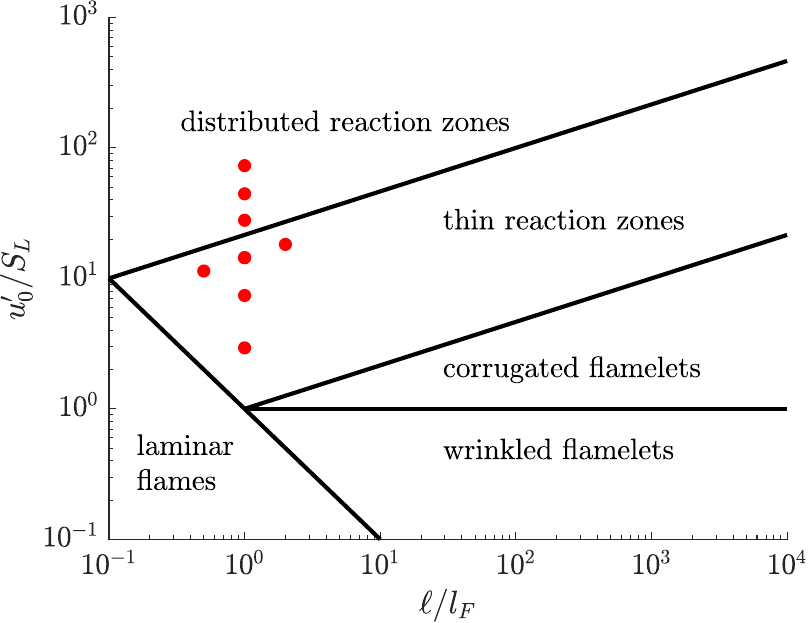}
    \caption{Turbulent premixed regime diagram. DNS test cases are marked with symbols.}
    \label{fig:regime}
\end{figure}

A summary of the simulation cases is presented in Table~\ref{tab:sim_cases} and shown in Fig.~\ref{fig:regime}. To conduct a comprehensive study of the effect of turbulence on unstable lean premixed hydrogen flames, simulations are carried out at six different Karlovitz numbers (A through F), and three different integral length scales (\chem{C_{0.5}}, \chem{C_1}, and \chem{C_2}). A fully laminar three-dimensional case is also included for reference (LAM).  As an additional point of comparison, cases A, B, and \chem{C_1} are also run with unity Lewis transport using the same forcing coefficient and domain, thus maintaining the same turbulence properties (e.g., turbulent kinetic energy, dissipation, etc.). However, as noted in Table~\ref{tab:laminar_parameters}, the laminar flame speed and thickness are different, and as such, the Karlovitz number and integral length scale ratio are different. These cases are denoted with a superscript $Le$ in Table~\ref{tab:sim_cases}.

The unburnt Karlovitz number is defined based on the ratio of flame and turbulence timescales:
\begin{equation}
    Ka_u = \frac{\tau_F}{\tau_\eta} = \frac{l_F}{S_L} \left(\frac{\varepsilon}{\nu_u}\right)^{1/2}
\end{equation}
where $\varepsilon$ is the dissipation, and $\nu_u$ is the kinematic viscosity. The flame timescale, $\tau_F$, is defined as $l_F/S_L$, and the Kolmogorov timescale, $\tau_\eta$, is $\sqrt{\nu_u/\varepsilon}$. The unburnt turbulent Reynolds number is defined as:
\begin{equation}
    Re_t = \frac{u' \ell}{\nu_u}
\end{equation}
where $u'$ is the turbulent intensity (rms velocity), and $\ell$ is the integral length scale.

The Karlovitz number, integral length scale, and turbulent intensity are often used to characterize the burning regime and are related through the following expression~\cite{peters2000turbulent}:
\begin{equation}
    \frac{u'}{S_L} = Ka^{2/3} \left(\frac{\ell}{l_F}\right)^{1/3}
\end{equation}
with the assumption that $S_Ll_F = \nu$. The relationship is conceptualized through the regime diagram, shown in Fig.~\ref{fig:regime}. The location of the DNS cases are marked with red symbols. The test cases span the thin reaction zone and distributed reaction zone regimes. Within the regimes, the tested Karlovitz numbers allow for a range of turbulent length scales, which synergize and compete with the cellular instabilities present in hydrogen combustion. The turbulence is sufficient to penetrate the flame and disrupt the flame structure~\cite{aspden2015turbulence} (e.g., species and source term profiles), which has a large effect on the global properties such as the flame speed. In this study, the integral length scale effects are tested at a fixed Karlovitz number, contrary to other studies which may instead choose to hold $u'/s_L$ constant~\cite{trivedi2022turbulence}. In this way, the flame is subject to an identical level of small-scale turbulence, and the effect of the integral length scale can be decoupled from the turbulence intensity and studied in isolation.

\begin{table*}[tbh] 
\caption{Parameters of the current simulations.}
\centerline{\begin{tabular}{l| c c c c c c c c c c c c}
Case           & A       & \chem{A^{Le}} & B       & \chem{B^{Le}} & \chem{C_{0.5}} & \chem{C_1} & \chem{C_1^{Le}} & \chem{C_2} & D        & E        & F & LAM \\ \hline
$Ka_u$         & $15$    & $4.4$         & $60$    & $17.5$        & $167$          & $167$      & $47.6$          & $167$      & $450$    & $900$    & $1900$   & $0$ \\
$Re_t$         & $24.13$ & $24.13$       & $60.79$ & $60.79$       & $47.05$        & $118.55$   & $118.55$        & $298.76$   & $229.29$ & $363.97$ & $598.97$ & $0$ \\
$L$ $[mm]$     & $4.03$  & $4.03$        & $4.03$  & $4.03$        & $2.01$         & $4.03$     & $4.03$          & $8.06$     & $4.03$   & $4.03$   & $4.03$   & $4.03$ \\
$\ell/l_F$     & $1$     & $1.72$        & $1$     &  $1.72$       & $0.5$          & $1$        & $1.72$          & $2$        & $1$      & $1$      & $1$      & $1$ \\
$A$ $[s^{-1}]$ & $314$   & $314$         & $792$   & $792$         & $2452$         & $1545$     & $1545$          & $973$      & $2987$   & $4742$   & $7803$   & $0$ 
\end{tabular}}
\label{tab:sim_cases}
\end{table*}

\section{Turbulence forcing}
\label{sec:forcing}
The flow configuration does not have any large-scale mean shear to generate the turbulence, such as in other studies~\cite{sankaran2015response}. Consequently, any injected turbulence would decay as it travels downstream. To prevent this, the flow must be forced across the domain in order to maintain a desired turbulence level. This forcing reproduces the effects of any mean shear not present in the simulation domain\cite{dhandapani2019using,dhandapani2020isotropic}.

The forcing vector, $\mathbf{f}$, in the momentum equation is the linear forcing of Lundgren~\cite{lundgren2003linearly,rosales2005linear} with the modification of Carroll and Blanquart~\cite{carroll2013proposed}. This is written as:
\begin{equation}
    \mathbf{f} = A \frac{k_0}{k(x,t)} \left(\rho \mathbf{u} - \overline{\rho \mathbf{u}}\left(x,t\right)\right)
\end{equation}
where the bar represents the planar Reynolds average:
\begin{equation}
    \overline{\phi} = \frac{1}{L^2} \int_0^L \int_0^L \phi \mathrm{dy dz}
\end{equation}
Here, $k_0$ represents the target turbulence kinetic energy (TKE), and
\begin{equation}
    k = \frac{1}{2} \left(\widetilde{\left(u''\right)} + \widetilde{\left(v''\right)^2} + \widetilde{\left(w''\right)}^2 \right)
\end{equation}
is planar Favre-averaged TKE. The Favre averaging is defined as:
\begin{equation}
    \widetilde{\phi} = \frac{\overline{\rho \phi}}{\overline{\rho}}
\end{equation}
with the corresponding fluctuating component
\begin{equation}
    \phi'' = \phi - \widetilde{\phi}
\end{equation}
This forcing technique has been used by many authors and validated in detail by Lapointe \textit{et al.}~\cite{lapointe2015differential} and Bobbitt \textit{et al.}~\cite{bobbitt2016vorticityTrans}. 

The forcing coefficient, $A$, directly controls the turbulence quantities of interest. For a given domain size (in the periodic directions), $L$, the integral length scale is approximately $\ell \approx 0.16 L$~\cite{bobbitt2016vorticityTrans}. Then, the nominal values for the rms perturbations, $u'_0$, TKE, $k_0$, dissipation, $\varepsilon_0$, and eddy turnover time, $\tau_0$, are written as~\cite{carroll2013proposed}:
\begin{align}
    u'_0 &= 3A\ell \\
    k_{0} &= \frac{27}{2}A^2 \ell^2 \\
    \varepsilon_0 &= 27 A^3 \ell^2 \\
    \tau_0 &= \frac{1}{2A}
\end{align}
The reported $Ka_u$ numbers are the nominal values given in terms of $\varepsilon_0$, which is directly controlled by $A$. To conduct a simulation at a desired $Ka_u$, the forcing coefficient is determined by:
\begin{equation}
    A = \left(\frac{1}{27}\frac{Ka_u^2 \nu_u S_L^2}{(0.16L)^2 l_F^2}\right)^{1/3}
\end{equation}
Turbulence is forced for the first 85\% of the domain. Then, the forcing coefficient is relaxed to 0 through a complementary error function in order to prevent negative velocities at the outlet.

To ensure realistic turbulence, an inflow file is generated from a simulation of homogenous isotropic turbulence (HIT). At each timestep, turbulent fluctuations from the HIT are superimposed onto a bulk mean velocity at the inlet. The fluctuations are initially smaller than the desired values and grow naturally to their nominal values with the forcing. 
\section{Overview of results}
\label{sec:overview}

In this section, we provide and overview of the effect of turbulence intensity on the turbulent flame brush. A qualitative discussion is provided first, followed by a quantitative discussion of the reaction zone broadening.

\subsection{Turbulent flame brush}
\label{subsec:flame_brush}

\begin{figure}[tbh]
    \centering
    \subfloat[From top to bottom: cases LAM, A, B, \chem{C_1}, D, E, and F]{
    \includegraphics[width=0.95\linewidth]{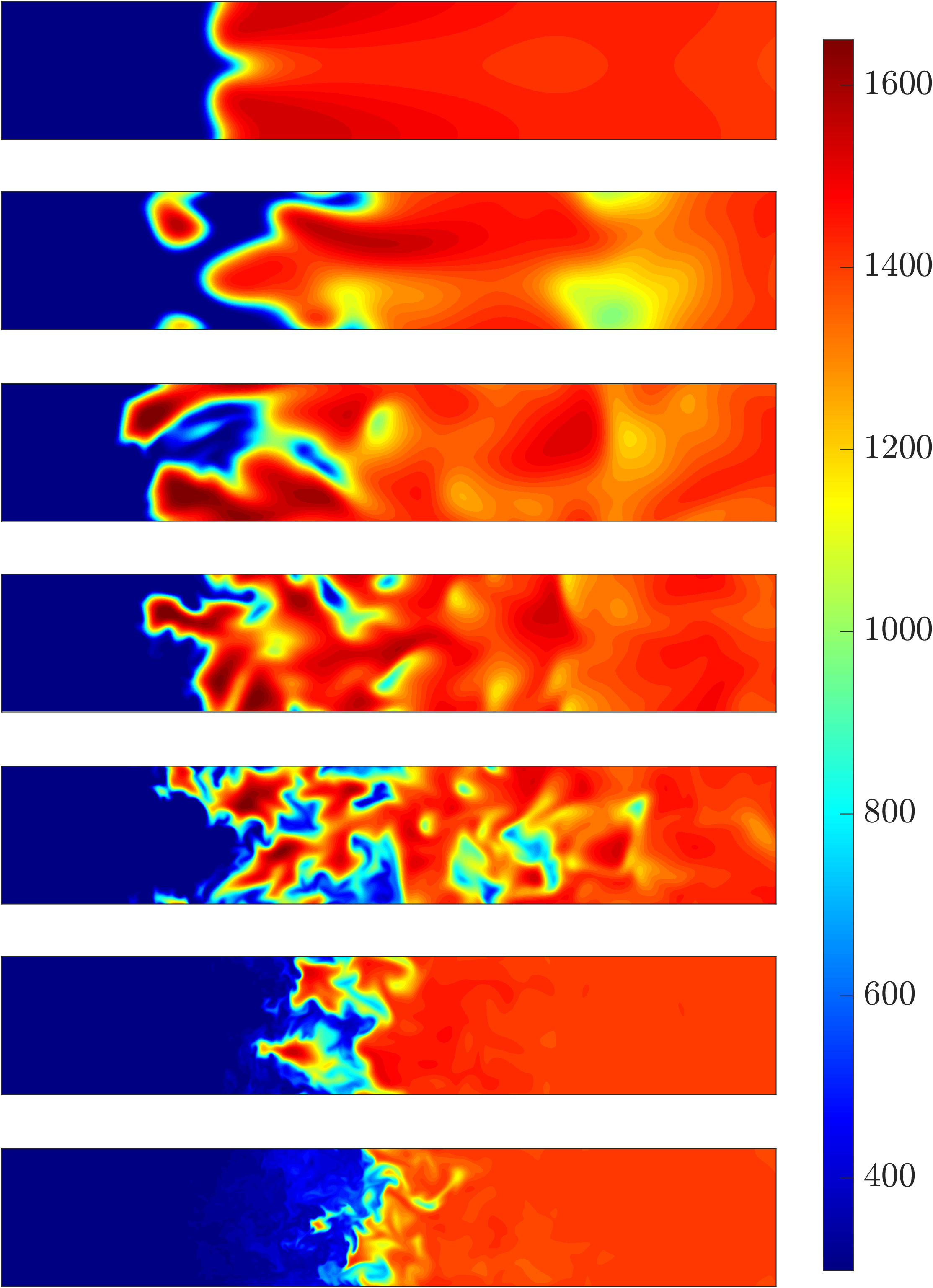}} \\
    \subfloat[From top to bottom: cases \chem{C_{0.5}}, \chem{C_1}, and \chem{C_2}]{
    \includegraphics[width=0.95\linewidth]{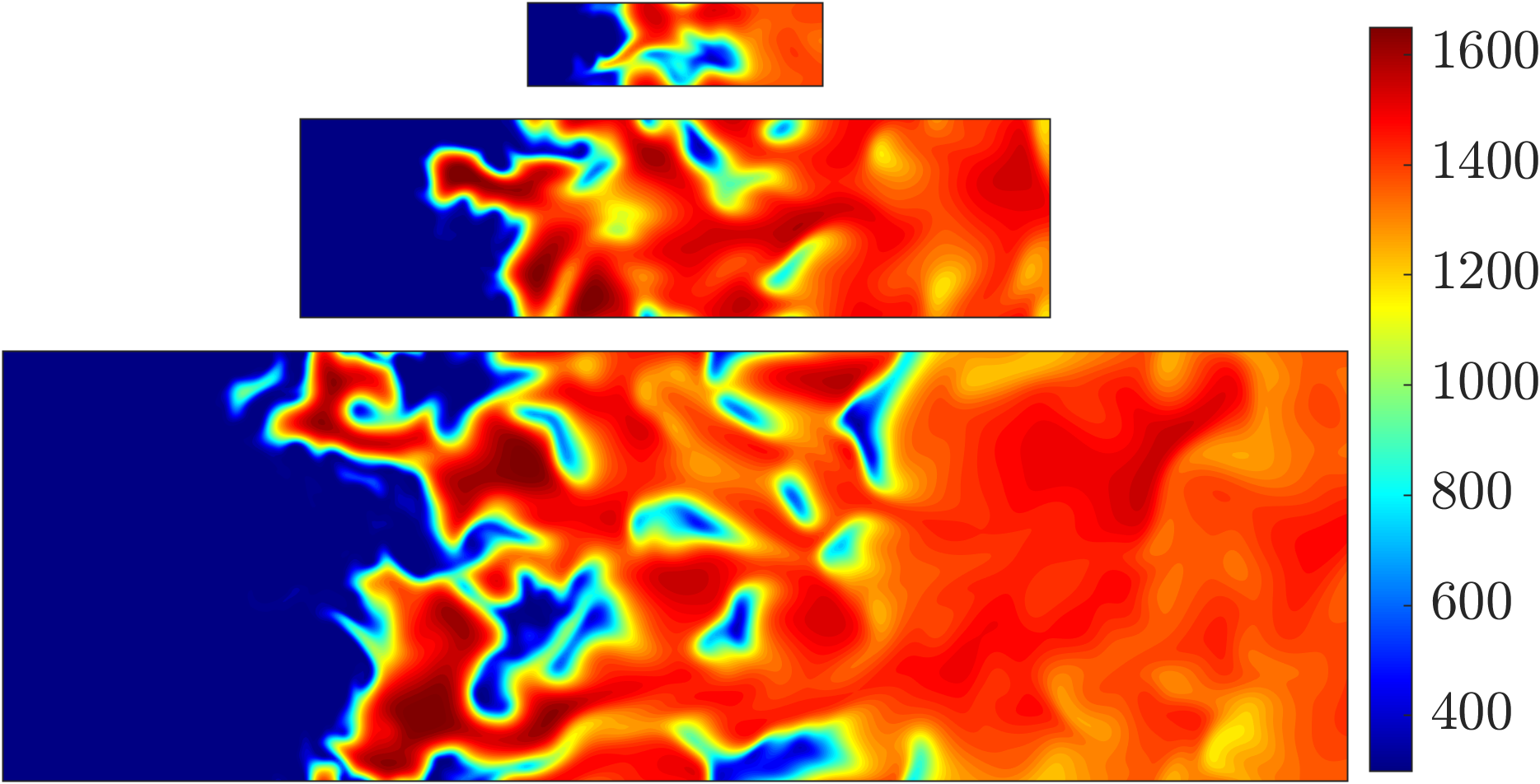}} 
    \caption{Two-dimensional slices of temperature field in the region of the flame. The figures are scaled to represent accurately the differences in physical length scales.}
    \label{fig:t_slice}
\end{figure}

Figure~\ref{fig:t_slice} shows two-dimensional slices of the temperature field for the tested cases. Due to the thermodiffusive instabilities, the 3D laminar flame (LAM) is not planar and exhibits cellular structures. The post flame temperature is seen to vary along the flame front. The curvature is defined to be positive when the center of curvature is located in the burnt mixture. The temperature is relatively higher in regions of positive curvature, and lower in regions of negative curvature~\cite{sanchez2014recent}. The preferential diffusion of \chem{H_2} concentrates the species into regions of positive curvature, creating a locally rich mixture which enhances the burning. The opposite is true for regions of negative curvature. The mismatch in local propagation speeds creates cellular structures which point toward the burnt mixture.  

Figure~\ref{fig:t_slice}a shows the effect of increasing the Karlovitz number at a fixed integral length scale ratio $\ell/l_F = 1$. For the lowest Karlovitz number, the flame exhibits a dominant structure with a continuous flame front. As the Karlovitz number increases, up until $Ka = 450$ (case D), the length scales of the structures at the flame front decrease. The flame front is increasingly disrupted by the turbulence, increasing the amount of small-scale features, leading to a significant increase in the turbulent flame area. The flame front appears more broken as pockets of burnt and unburnt gases are mixed by the turbulence. Due to the thermodiffusive instabilities, the temperature field in the burnt mixture is highly inhomogeneous. At the highest Karlovitz number, $Ka = 1900$ (case F), the turbulent flame brush becomes significantly shorter, and the mixing is smoother as there are less disconnected pockets of fluid. Qualitatively, flame F appears similar to thermodiffusively stable high Karlovitz number hydrocarbon flames~\cite{lapointe2015differential}. 

Figure~\ref{fig:t_slice}b shows the effect of changing the integral length scale at a fixed Karlovitz number. As the integral length scale is increased, the flame presents significantly more disconnected regions as pockets of unburnt gases are mixed into the burnt mixture, increasing the turbulent flame area. Despite the smaller domain size restricting the volume that the flame can grow in, the structures at the flame front are qualitatively of the same size. It is important to note that for all three cases, the Karlovitz number was kept constant by maintaining the same amount of dissipation. Although the large scale turbulence that the flames are subject to is changing, the flames are still subject to the same small scale turbulence. Thus, it is reasonable that the flames exhibit similar small-scale structures.

\subsection{Reaction zone broadening}

As the Karlovitz number is increased, the role played by turbulent mixing is enhanced. This can lead to reaction zone broadening and is the first step towards distributed burning~\cite{driscoll2020premixed}.

\begin{figure}[tbh]
    \centering
    \includegraphics[width=1\linewidth]
    {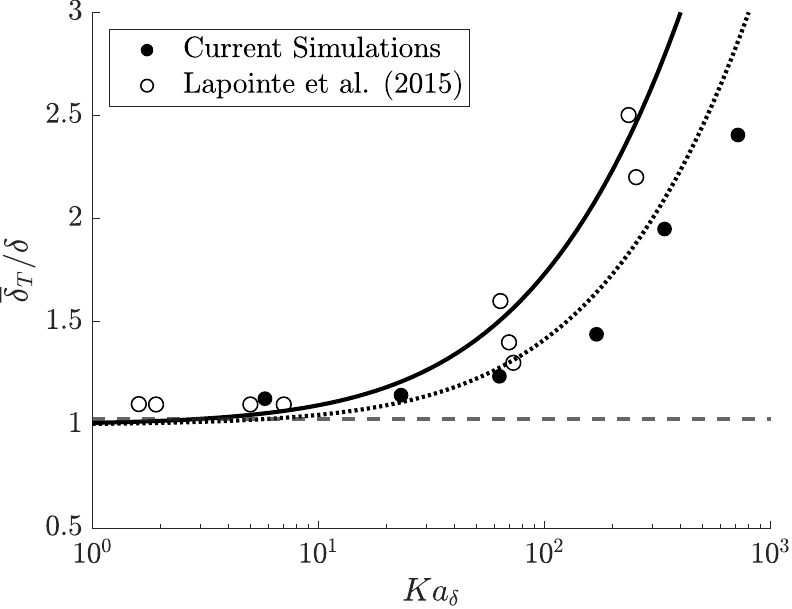}
    \caption{Effective reaction zone thickness normalized by the laminar reaction zone thickness. The dashed line is the value obtained from case LAM. Solid ($a=0.02$) and dotted ($a=0.01$) lines represent the $Ka_\delta$ model (Eq.~\eqref{eq:delta_model}). Data from Lapointe \textit{et al.}~\cite{lapointe2015differential} for n-heptane/air flames is also included.}
    \label{fig:delta}
\end{figure}

The laminar reaction zone thickness, $\delta$, is defined as the distance over which the fuel consumption rate exceeds 5\% of its maximum value. For the turbulent flames, the effective mean reaction zone thickness is calculated in a similar fashion to Lapointe \textit{et al.}~\cite{lapointe2015differential} using the ratio of the volume where the fuel consumption rate exceeds 5\% of the maximum laminar value and the turbulent flame area (defined later):
\begin{equation}
\label{eq:thickness}
    \overline{\delta_T} = \frac{\overline{V}\left(\dot{\omega}_F > 0.05 \dot{\omega}_{F,\mathrm{max}}^{1D}\right)}{\overline{A_T}}
\end{equation}
The effective reaction zone thickness normalized by the laminar reaction zone thickness is shown in Fig.~\ref{fig:delta}. To eliminate the effect of different ambient temperatures when comparing results with previous simulations~\cite{lapointe2015differential}, the reaction zone thickness is plotted against the reaction zone Karlovitz number:
\begin{equation}
    Ka_{\delta} = \frac{\delta^2}{\eta_\delta^2}
\end{equation}
where $\eta_\delta$ is the Kolmogorov calculated using $\nu$ at $C_{peak}$. Note the 3D laminar case has a reaction zone thickness almost identical to the one-dimensional flame ($\overline{\delta_T}/\delta \approx 1.03$). It is interesting to contrast these results with those from Fig.~\ref{fig:t_slice}. As the Karlovitz number is increased from case A to D, the overall turbulent flame brush becomes larger, indicating that the flame is occupying more volume.  However, locally, there does not appear to be much broadening of the reaction zone. When the Karlovitz number is increased further from case D to case F, flame broadening becomes more noticeable, while the turbulent flame brush decreases.  For the highest Karlovitz number case, the reaction zone is approximately 2.4 times that of the laminar reaction zone thickness. 

Data from Lapointe \textit{et al.}~\cite{lapointe2015differential} for n-heptane/air flames is also plotted and shows excellent qualitative agreement with the present results.  The similarity between the results suggest that chemistry has a negligible impact on the broadening of the reaction zone.  Attributing it to the enhanced diffusivity due to turbulence, Lapointe \textit{et al.}~\cite{lapointe2015differential} proposed the following scaling:
\begin{equation}
    \frac{\overline{\delta_T}}{\delta} \propto \sqrt{1+aKa_\delta}
    \label{eq:delta_model}
\end{equation}
where $a$ is a proportionality constant which is used to fit the data. The proposed scaling is able to predict the increase in reaction zone thickness with Karlovitz number.  For the present present lean hydrogen/air flames, a lower coefficient is obtained ($a=0.01$) than was extracted ($a=0.02$) for the {\it n}-heptane/air flames~\cite{lapointe2015differential}.

\section{Global properties}
\label{sec:glob_loc}

The burning efficiency model of Eq.~\eqref{eq:efficiency} aims to explain the relationship between two global quantities, the turbulent flame speed and area. In this section, a detailed study of the turbulent flame speed and area is conducted. Then, an expression for the burning efficiency is proposed.

\subsection{Turbulent flame speed}

\begin{figure}[tbh]
    \centering
    \subfloat[Instantaneous]{
    \includegraphics[width=0.95\linewidth]{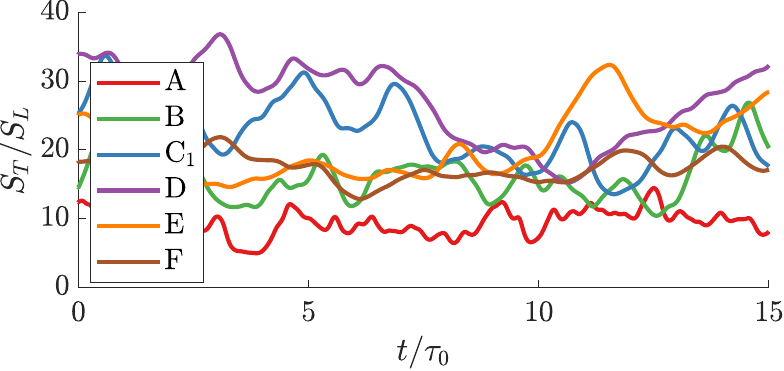}} \\
    \subfloat[Mean]{
    \includegraphics[width=0.95\linewidth]{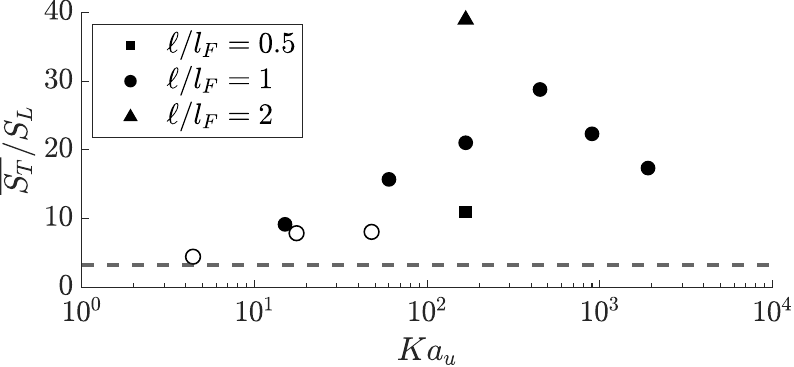}} 
    \caption{Instantaneous (top) and mean (bottom) normalized turbulent flame speed.}
    \label{fig:st_ka_mean}
\end{figure}

From integrating the continuity and water mass fraction equations, the turbulent flame speed is defined as the volume integral of the progress variable production rate:
\begin{equation}
\label{eq:ST}
    S_T = \frac{1}{\rho_u C_b A} \int_V \dot{\omega}_{C} \mathrm{dV}
\end{equation}
where $A = L\times L$ is the cross sectional area. An example of the temporal evolution of the turbulent flame speed is shown in Fig.~\ref{fig:st_ka_mean}a for the different Karlovitz numbers. 
The mean turbulent flame speeds are shown in Fig.~\ref{fig:st_ka_mean}b as a function of the Karlovitz number. The turbulent flame speed increases with the Karlovitz number up to Case D, after which the flame speed decreases. A similar trend was noted by Aspden \textit{et al.}~\cite{aspden2019towards}. 

To explain the evolution of the turbulent flame speed, it is insightful to rewrite the integral in Eq.~\eqref{eq:ST} as a surface integral along an isocontour of $C$ and an integral in the normal direction:
\begin{equation}
\label{eq:src_vol_integral}
    \int_V \dot{\omega}_C \mathrm{dV} = \int_n \left( \iint_{A(C)} \dot{\omega}_C \mathrm{dA} \right) \mathrm{dn}
\end{equation}
Taking $\mathrm{dn} = \mathrm{dC}/\vert \nabla C \vert$, Eq.~\eqref{eq:src_vol_integral} can be rewritten as:
\begin{equation}
    \int_V \dot{\omega}_C \mathrm{dV} = \int_C \iint_{A(C)} \frac{\dot{\omega}_C}{\vert \nabla C\vert} \mathrm{dAdC}
\end{equation}
and the turbulent flame speed can be written as:
\begin{equation}
\label{eq:int_final}
    S_T = \frac{1}{\rho_uC_bA} \int_C \left \langle \frac{\dot{\omega}_C}{\vert \nabla C \vert} \bigg \vert C\right \rangle A(C) \mathrm{dC}
\end{equation}
The turbulent flame speed can be interpreted as a convolution of the area of the flame isosurface, $A(C)$, with the gradient weighted conditional mean source term. 

Turbulence can thus impact the flame speed through one of two main pathways. The first pathway is through local effects induced by the turbulence on the mean source term, $\langle \frac{\dot{\omega}_C}{\vert \nabla C \vert} \vert C \rangle$. As the small scale turbulence penetrates the flame, it alters the flame structure through enhanced mixing and diffusion. The second pathway is through the  area of flame isosurfaces, $A(C)$. As the larger scale turbulence perturbs the flame front, it wrinkles the flame, altering the curvature, $\kappa$, and necessarily increasing the flame area. These effects are detailed in the following subsections.

\subsection{Area of flame isosurfaces}

Previous studies have identified the instantaneous flame front as the isosurface of either $T_{peak}$ or $C_{peak}$~\cite{lapointe2016fuel,day2009turbulence}. The flame area is then subsequently calculated as the area of that isosurface. However, the presence of thermodiffusive instabilities in the current simulations necessitate some nuances on the definition of the flame area.

\begin{figure}[tbh]
    \centering
    \includegraphics[width=1\linewidth]
    {{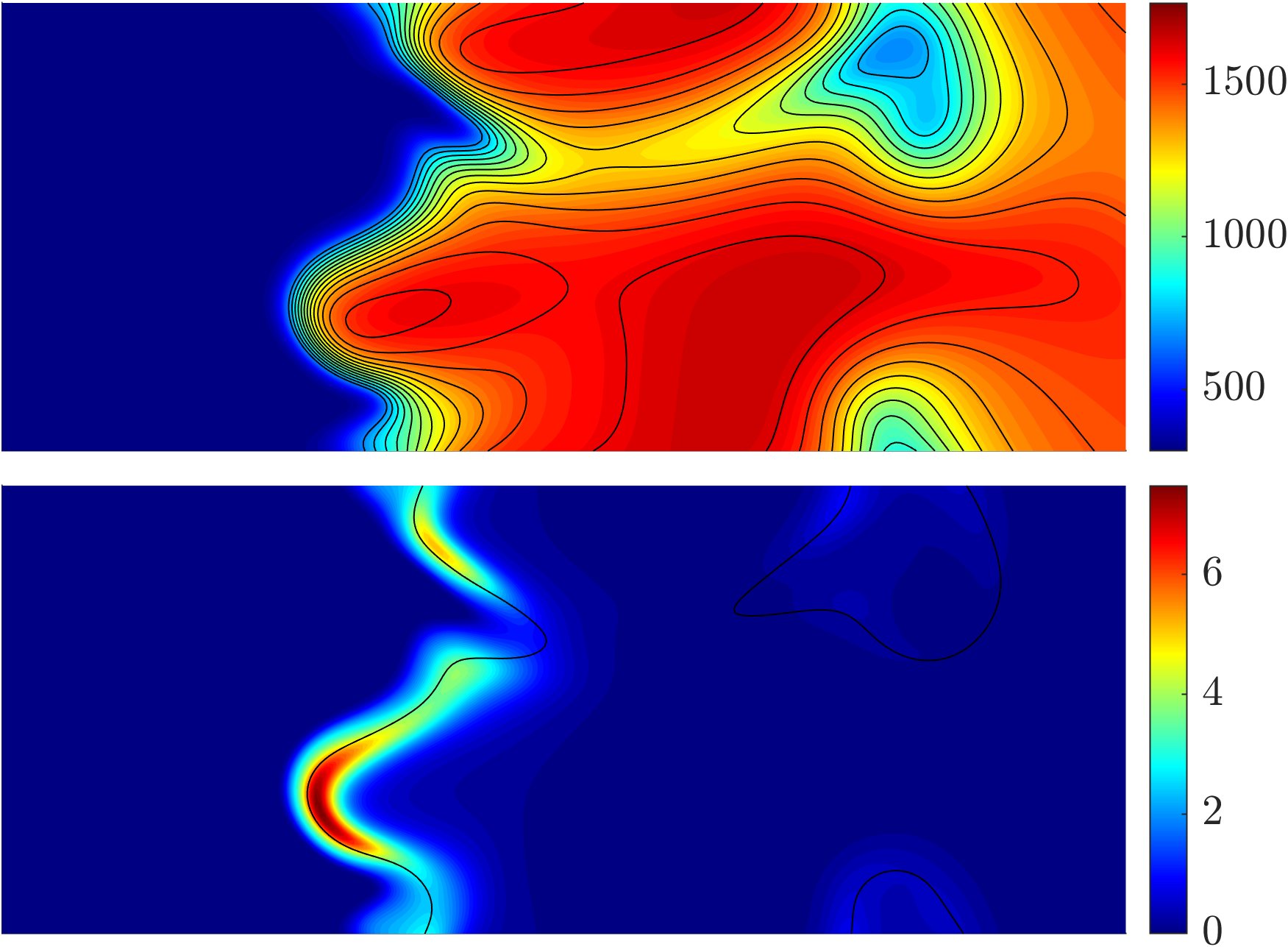}}
    \caption{Temperature, $T$ (top), and normalized source term, $\dot{\omega}_C/\dot{\omega}_{C,max}^{1D}$ (bottom) for case A. Black isolines on the temperature represent 0.4 to 1.6 times $C_{peak}$ in increments of 0.1. The black isoline on the source term is at $C_{peak}$.}
    \label{fig:T_SRCwithcontours}
\end{figure}

Fundamentally, the flame front is described as the location where the reactions are occurring. Figure~\ref{fig:T_SRCwithcontours} shows the temperature and the normalized source term with  isocontours of $C$ from case A. From the upper plot, it can be seen that the isocontours are irregularly spaced, indicating that the flame area varies  depending on the value of $C$ chosen as the isocontour value. From the lower plot, along the isocountour of $C_{peak}$, the source term varies with the curvature of the iscountour. In regions of positive curvature (center of curvature in burnt mixture), the source term is about 7.5 times larger than the maximum source term from the one-dimensional flame. The curvature concentrates the differential diffusion into these regions, and as a result, the mixture becomes locally richer and the burning is significantly enhanced.  Hydrogen diffuses out of regions of negative curvature, creating locally leaner mixtures, and the source term reduces. Interestingly, the source term plot shows a pocket of non-burning low temperature mixture which has been engulfed into the burnt mixture and is captured by the isosurface of $C_{peak}$. This artificially increases the flame area, and the effects can become more pronounced as the flow fields become more complex with higher levels of turbulence, as shown in Fig.~\ref{fig:t_slice}. 

\begin{figure}[tbh]
    \centering
    \includegraphics[width=1\linewidth]
    {{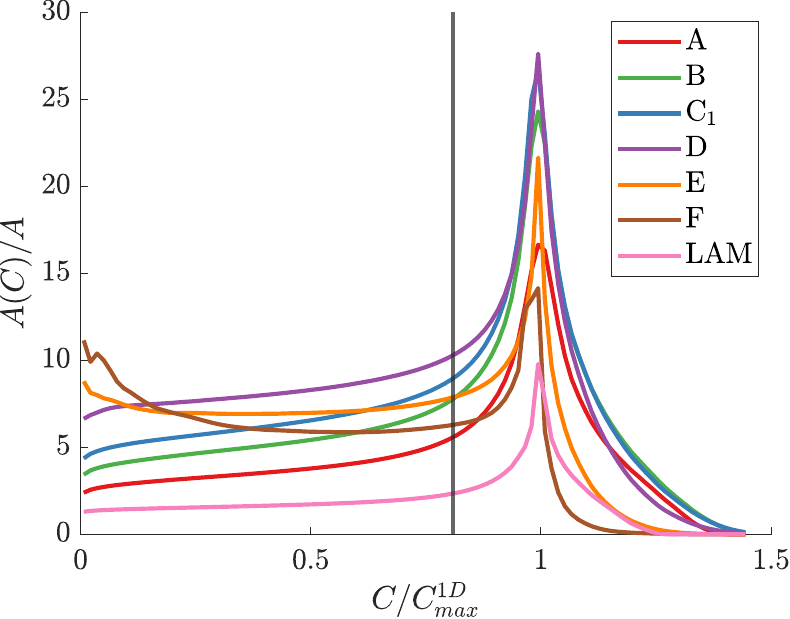}}
    \caption{Normalized area of flame isosurfaces as a function of normalized progress variable for flames at $\ell/l_F = 1$. The vertical line denotes the location of $C_{peak}$.}
    \label{fig:AC}
\end{figure}

To illustrate the effect of the isosurface value selection, Fig.~\ref{fig:AC} shows the normalized area of isosurfaces as a function of C, $A(C)/A$, for the various Karlovitz number flames.  The isosurfaces are created using the classical marching cubes method~\cite{loresnsen1987marching}.  The $A(C)$ profiles may be decomposed into two regions.  At lower values of the progress variable (for $C<C_{peak}$), $A(C)$ increases modestly with progress variable.  At a fixed $C$, it increases monotonically with the Karlovitz number from case LAM to case D.  Such increase is expected: more intense turbulence means more flame wrinkling and hence larger isosurface area.  At larger values of the progress variable (for $C>C_{peak}$), $A(C)$ increases sharply with progress variable before reaching a large spike at $C=C_{max}^{1D}$, with a subsequent drop-off to 0.  The spike is indicative of the strong post-flame inhomogeneities present due to differential diffusion effects. In both regions, as the Karlovitz number is further increased (cases E and F), the differential diffusion effects are suppressed. This translates into overall flatter $A(C)$ profiles, with smaller spikes and quicker decay to zero in superadiabatic regions ($C>C_{max}^{1D}$).

\begin{figure*}[tbh]
    \centering
    \subfloat[Effect of integral length scale]{
    \includegraphics[width=0.33\textwidth]{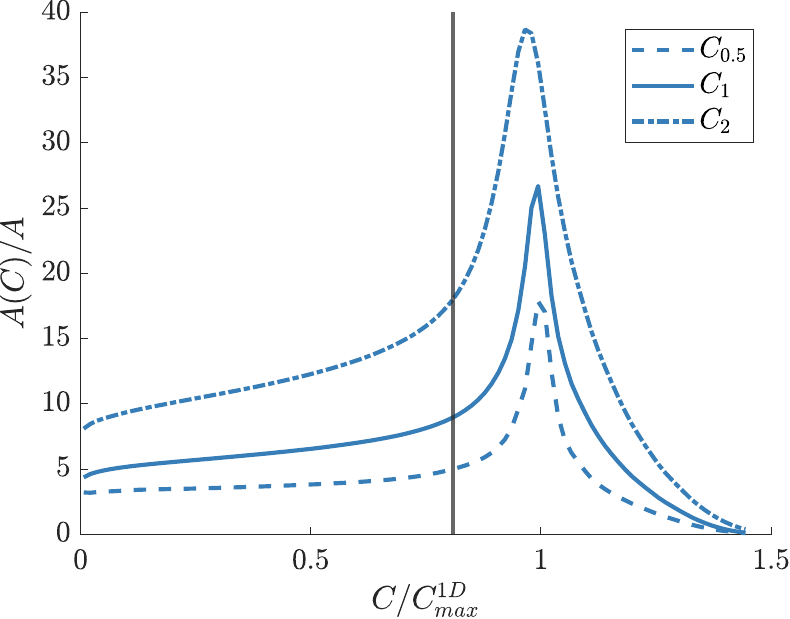}} 
    \subfloat[Scaled by integral length scale ratio]{
    \includegraphics[width=0.33\textwidth]{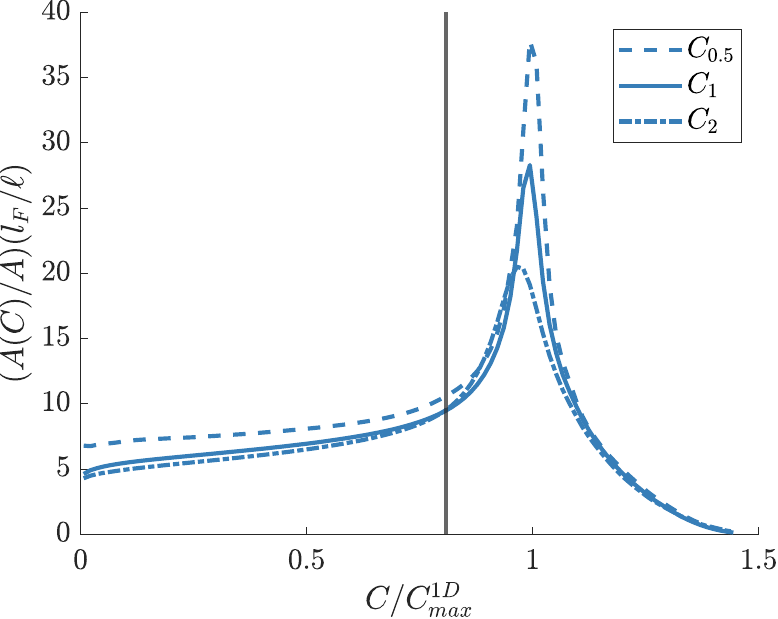}} 
    \subfloat[A snapshot of case \chem{C_2} (line), compared to the flame isosurface area for each quarter of the domain (markers).]{
    \includegraphics[width=0.33\textwidth]{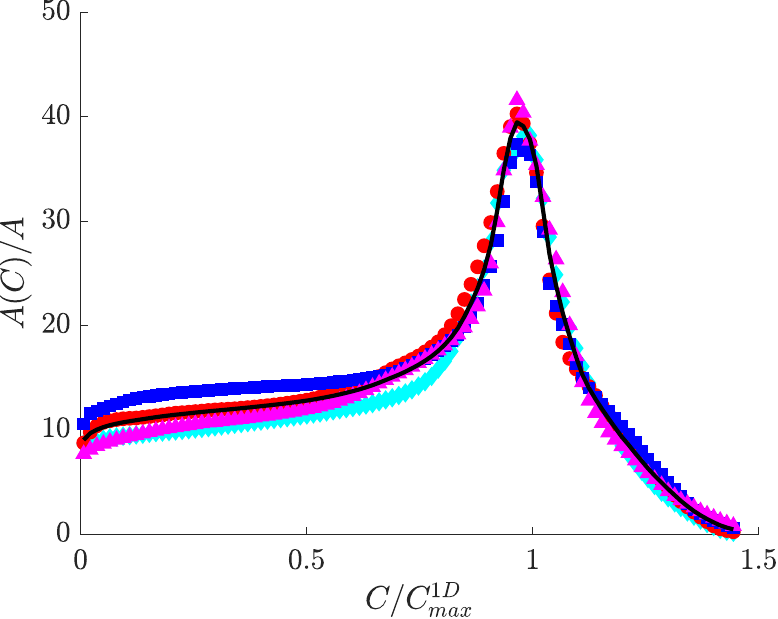}} 
    \caption{Effect of integral length scale and domain area on the flame isosurface area.}
    \label{fig:ACint}
\end{figure*}

The influence of the integral length scale on the flame area is shown in Fig.~\ref{fig:ACint}. Figure~\ref{fig:ACint}a shows the effect of the integral length scale ratio on the flame isosurface area for cases $C_x$. As $\ell/l_F$ is increased, the ratio $A(C)/A$ increases systematically.  The larger integral length scales introduce more large-scale wrinkling. A linear dependence was observed for turbulent hydrocarbon flames by Lapointe~\cite{lapointe2016simulation}.  To investigate this dependence further, Fig.~\ref{fig:ACint}b shows $A(C)/A$ scaled by $\ell/l_F$. The profiles are almost collapsed, indicating that the $A(C)/A$ scales almost linearly with the integral length scale ratio.

n the present simulations, the domain size and integral length scale are inseparably linked since $\ell \approx 0.16L$~\cite{bobbitt2016vorticityTrans}. As pointed out by Aspden~\cite{aspden2019towards}, it is unclear if the domain size or the integral length scale are the cause of the linear of the ratio $A(C)/A$ .  To elucidate their individual contributions, a single snapshot of case $C_2$ is decomposed into four quadrants.  Each quadrant is characterized by the same integral length scale as the original full domain, but has a reduced domain width by a factor of 2.  The normalized flame isosurface areas for all four quadrants and the full domain are shown in Fig.~\ref{fig:ACint}c. The overlapping profiles indicate that the ratio $A(C)/A$ is independent of the domain width.  The domain width does impact the flame isosurface area $A(C)$ through the cross section area $A=L^2$, but not their ratio.

\subsection{Turbulent flame area}

Given the large variations in $A(C)$, defining a single $A_T$ value for a given flame presents some arbitrariness, and hence different definitions may be proposed.

\begin{figure}[tbh]
    \centering
    \includegraphics[width=1\linewidth]
    {{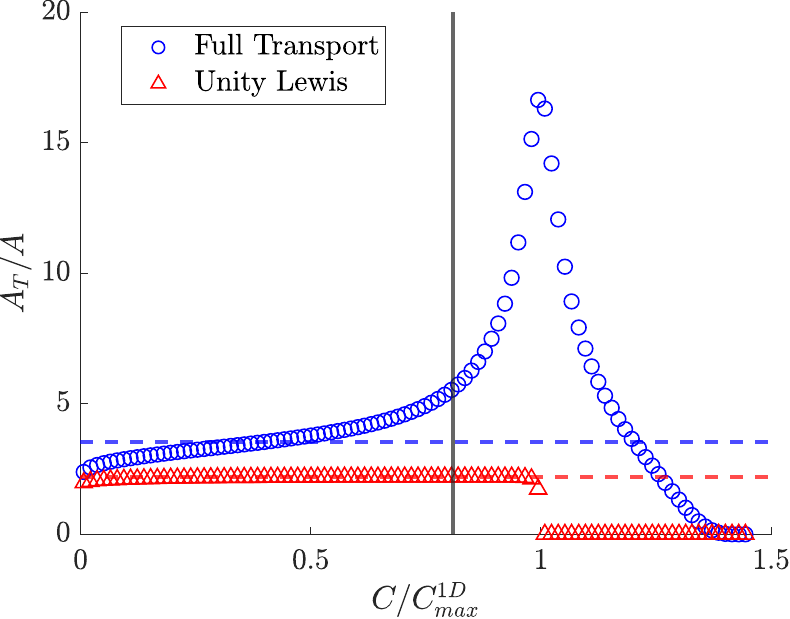}}
    \caption{Area of flame isosurfaces as a function of progress variable for flame A with full transport (circles) and unity Lewis transport (triangles).  The dashed lines correspond to the mean turbulent flame areas calculated using Eq.~\eqref{eq:AT}.  The vertical line denotes the location of $C_{peak}$.}
    \label{fig:ka15_AT}
\end{figure}

Inspired by Eq.~\eqref{eq:int_final}, a natural definition is to use a source term-weighted average flame area:
\begin{equation}
\label{eq:AT}
    A_T \equiv \frac{\bigintss \left \langle \frac{\dot{\omega}_C}{\vert \nabla C \vert} \big \vert C \right \rangle A(C) \mathrm{dC}}{\bigintss \left \langle \frac{\dot{\omega}_C}{\vert \nabla C\vert} \big \vert C \right \rangle \mathrm{dC}} 
\end{equation}
The weighting gives more importance to areas with enhanced burning at the flame front, and significantly reduces the contribution of non-burning regions, such as those shown in Fig.~\ref{fig:T_SRCwithcontours}. Figure~\ref{fig:ka15_AT} shows the area of flame isosurfaces for case A.  For the stable unity Lewis number case, the flame area is approximately constant regardless of the chosen isosurface value. This constant value is well reproduced by the source term-weighted average flame area.  For the full transport case, the turbulent flame area traditionally calculated as the area of the isosurface of $C_{peak}$, $A(C_{peak})$, is approximately 1.5 times larger than the mean area from Eq.~\eqref{eq:AT}. For this given case, the turbulent flame area calculated using Eq.~\eqref{eq:AT} is the same as the area of the isosurface $C/C^{1D}_{max} \approx 0.4$.

\begin{figure}[tbh]
    \centering
    \subfloat[Effects of Karlovitz number]{
    \includegraphics[width=0.8\linewidth]{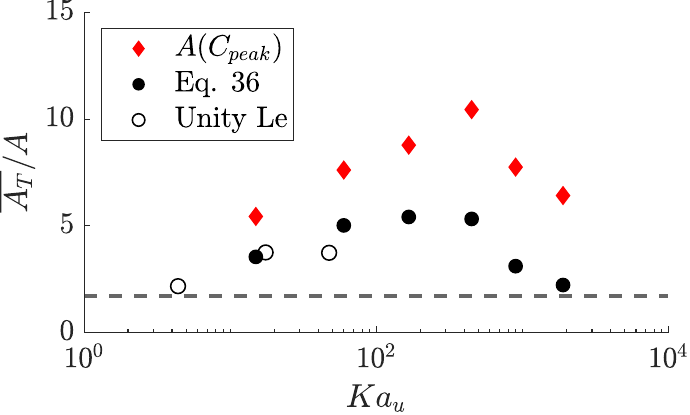}} \\
    \subfloat[Effects of integral length scale ratio]{
    \includegraphics[width=0.8\linewidth]{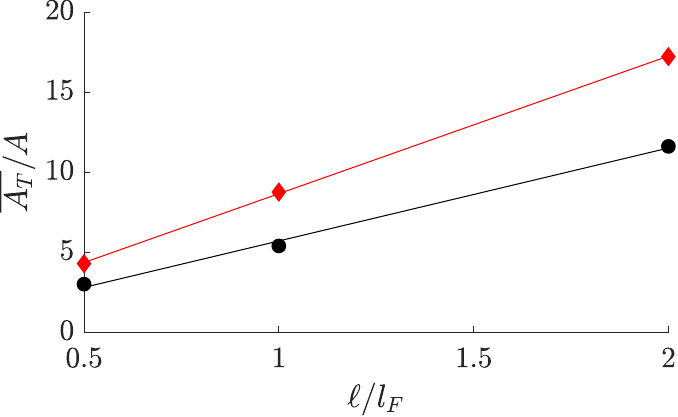}} 
    \caption{Impact of the Karlovitz number (top) and integral length scale ratio (bottom) on the mean normalized turbulent flame area. The open symbols represent flames calculated with unity Lewis numbers, and the dashed lines are values obtained from case LAM.}
    \label{fig:at_mean}
\end{figure}

The influences of the Karlovitz number and integral length on the mean turbulent flame areas are shown in Fig.~\ref{fig:at_mean}. The area exhibits a minor dependence on the Karlovitz number, increasing from about 2.0 for the LAM case to about 5.0 for case D.  The dependence on the integral length scale is stronger, with an almost linear increase with the ratio $\ell/l_F$, consistent with the results shown in Fig.~\ref{fig:ACint}. In hydrocarbon flames, Lapointe~\cite{lapointe2016simulation} also found a linear dependence of $A_T/A$ on $\ell/l_F$, up to an integral length scale ratio of 4. 
One may note that the turbulent flame area calculated using Eq.~\eqref{eq:AT} is systematically lower than the area of the isosurface of $C_{peak}$. However, the trends remain the same.

The decrease of $A_T/A$ at the highest Karlovitz numbers is hypothesized to be attributed to two main sources. The first is that the flame becomes too broad, and the domain is too small to accommodate the associated large-scale wrinkling (see section~\ref{subsec:flame_brush}). The second is that the thermodiffusive instabilities are suppressed as the effective Lewis number approaches unity, thereby reducing the flame area which was being enhanced.

\subsection{Burning efficiency}

The turbulent flame speed and area are related to each other through the burning efficiency parameter from Eq.~\eqref{eq:efficiency}. As such, the definition of the turbulent flame area is intrinsically linked to that of the burning efficiency.

\begin{figure}[tbh]
    \centering
    \includegraphics[width=1\linewidth]
    {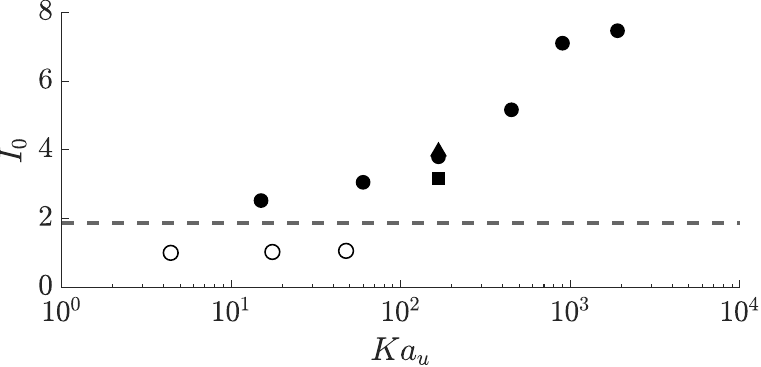}
    \caption{Burning efficiency as calculated by Eq.~\eqref{eq:i0} versus Karlovitz number. The dashed line is obtained from case LAM. Open symbols represent unity Lewis solutions.}
    \label{fig:i0_ka}
\end{figure}

Taking the expressions for $S_T$ and $A_T$ from Eqs.~\eqref{eq:int_final} and ~\eqref{eq:AT}, then $S_T/S_L$ can be written as:
\begin{equation}
    \frac{S_T}{S_L} = \frac{\bigintss \left \langle \frac{\dot{\omega}_C}{\vert \nabla C \vert} \big \vert C \right \rangle \mathrm{dC}}{\bigintss  \frac{\dot{\omega}_C^{lam}}{\vert \nabla C \vert^{lam}}  \mathrm{dC}} \frac{A_T}{A} 
\end{equation}
This equation is mathematically equivalent to Eq.~\eqref{eq:efficiency} with 
\begin{equation}
\label{eq:i0}
    I_0 \equiv \frac{\bigintss \left \langle \frac{\dot{\omega}_C}{\vert \nabla C \vert} \big \vert C \right \rangle \mathrm{dC}}{\bigintss  \frac{\dot{\omega}_C^{lam}}{\vert \nabla C \vert^{lam}}  \mathrm{dC}}
\end{equation}
The expression for $I_0$ is a generalization of the one proposed by Lapointe and Blanquart~\cite{lapointe2016fuel}, as it does not rely on the key assumption that the source term scales with its value at $C_{peak}$. The calculated values of $I_0$ are plotted in Fig.~\ref{fig:i0_ka}. There are three key observations.  

First, at a fixed Karlovitz number, the burning efficiency appears to be independent of the integral length scale, as the values for $\ell/l_F=1$ and $2$ are virtually identical. The lower burning efficiency of the $\ell/l_F = 0.5$ case deserves more attention. An integral length scale less than the flame thickness means a sub-optimal amount of turbulence penetrating the flame.  With less turbulence impacting the flame, the burning efficiency is lower.

Second, the burning efficiency increases monotonically with Karlovitz number.  At low turbulence intensities, the burning efficiency asymptotes to the value from the LAM flame. Notably, due to the thermodiffusive instabilities, the burning efficiency of the laminar case is almost 2, indicating that the instabilities play a significant role in the burning efficiency enhancement.  It should be noted that the increase with Karlovitz number continues even after case D, where the turbulent flame speed and area begin to decrease. At the highest Karlovitz numbers, differential diffusion effects are virtually suppressed. Consequently, the relevant laminar flame speed would that of a unity Lewis number flame, which is larger than that with full transport (see Table~\ref{tab:laminar_parameters}). Since the current results are calculated based on the nominal $S_L$ for full transport, the values of $I_0$ may be inflated at higher $Ka_u$. 

These trends build upon the results obtained for the stable hydrocarbon flames of Lapointe and Blanquart~\cite{lapointe2016fuel}. Their burning efficiencies were found to be very close to unity, and almost constant for all of their tested Karlovitz numbers. Notably, because their flames were thermodiffusively stable, the difference in the burning efficiencies between the mixture average and unity Lewis cases is much less. 

\section{Local flame response}
\label{sec:local}

As described in Eq.~\eqref{eq:i0}, the burning efficiency is dictated by the local response of the flame to the imposed turbulence. In this section, we explore the interplay between chemical source term, turbulence, and flame curvature.

\subsection{Chemical source term}

\begin{figure}[tbh]
    \centering
    \subfloat[Effect of integral length scale]{
    \includegraphics[width=1\linewidth]{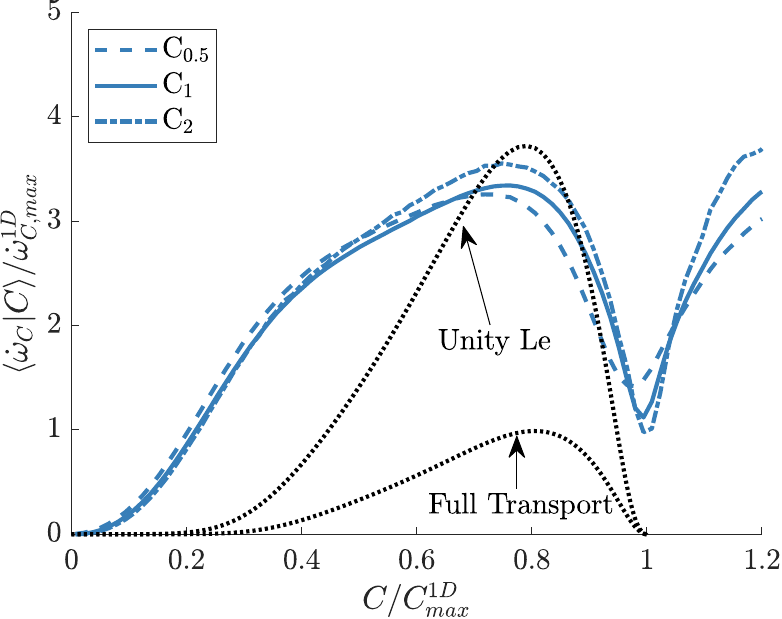}} \\
    \subfloat[Effect of Karlovitz number]{
    \includegraphics[width=1\linewidth]{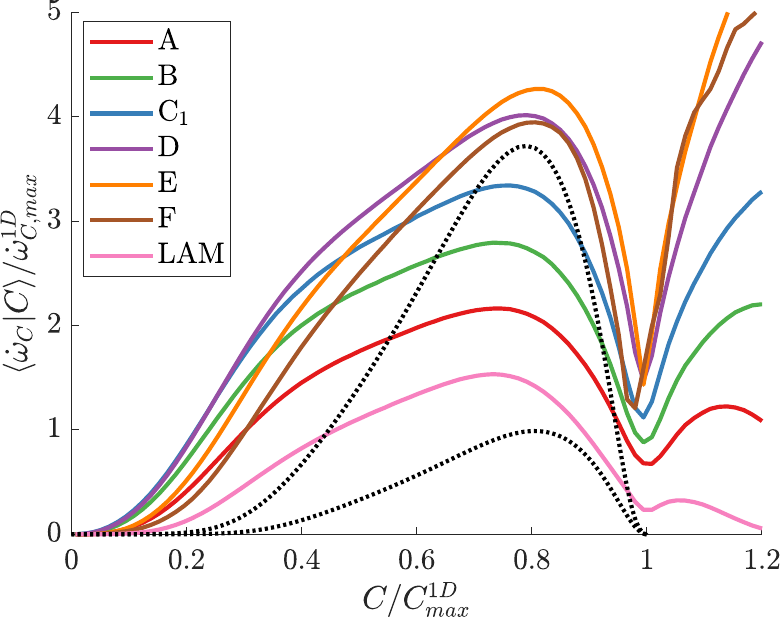}} 
    \caption{Effect of the integral length scale (top) and Karlovitz number (bottom) on the conditional mean source term profiles. The dotted lines indicate profiles from limiting one-dimensional flames.}
    \label{fig:src_c}
\end{figure}

The conditional mean progress variable source term profiles are plotted in Fig.~\ref{fig:src_c}. Consistent with the previous observation related to the burning efficiency, at a given Karlovitz number, the source term profiles in Fig.~\ref{fig:src_c}a show only a weak dependence on the integral length scale. The profiles show strong agreement until the peak, where the peak value shows a slight increase with increasing integral length scale. However, the differences are minor in comparison to the increase in flame area with increasing integral length scale. Non-zero source terms beyond the nominal maximum $C_{max}^{1D}$ value correspond to superadiabatic mixtures.

Regardless of the Karlovitz number, the shape of the profiles (Fig.~\ref{fig:src_c}b) is changed significantly, even for the laminar case. Before the peak source terms, the three-dimensional mean profiles exhibit a bump (around $C\approx0.35$) which is not present for the one-dimensional profiles.  For all cases, the area under the curve is significantly larger than for the one-dimensional profiles. The thermodiffusive instabilities spread the burning across a wider range of $C$ values. Although the magnitude of the source terms is different, they share roughly the same shape until case D. After case D, the profiles start to resemble the unity Lewis profile more closely.

\begin{figure}[tbh]
    \centering
    \subfloat[Maximum value]{
    \includegraphics[width=1\linewidth]{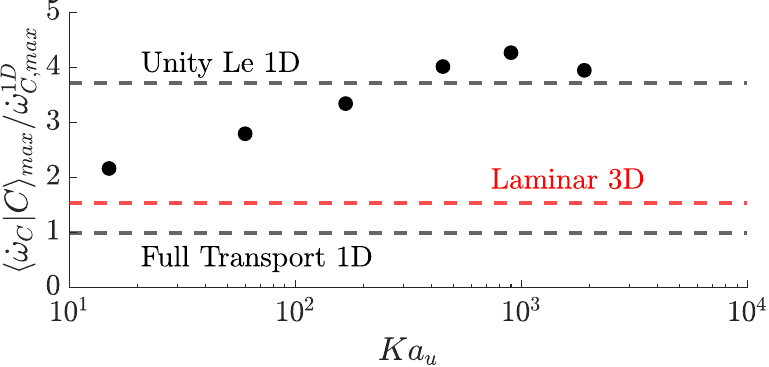}} \\
    \subfloat[Location of maximum]{
    \includegraphics[width=1\linewidth]{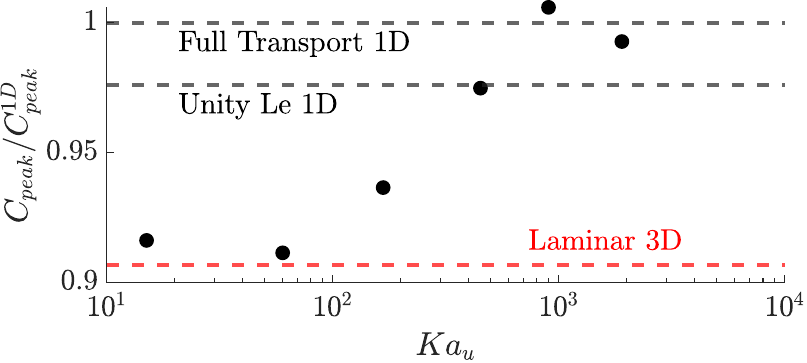}} 
    \caption{Peak normalized conditional mean progress variable source term (top) and its location in progress variable space (bottom).}
    \label{fig:src_max_cpeak}
\end{figure}

To quantify this evolution, Fig.~\ref{fig:src_max_cpeak} shows the evolution of the peak source term and its location versus Karlovitz number. The magnitude is found to increase until case D, where it reaches a steady value, slightly above that of the unity Lewis number flamelet. Similarly, the location of the peak source term shifts to reach the value of the one-dimensional unity Lewis number flame.  This asymptoting to the unity Lewis number values at high Karlovitz numbers is the result of the gradual disappearance of differential diffusion effects and was also observed in hydrocarbon flames~\cite{lapointe2015differential}.

\subsection{Propagating surface vs material surface}

To provide insight into the evolution of the flame front, we follow the analysis of Yeung et al.~\cite{yeung1990straining} which consider a flame front as an infinitely thin propagating surface with an intrinsic propagation speed.  As turbulence is introduced, a competition arises between self propagation and flow-induced motion. Initially, the flame front behaves as a propagating surface and can respond to the turbulence-induced curvature.  However, once turbulence is increased beyond a certain point, the flame acts as a material surface, convected by the flow and unable to control its shape.  At this point, the geometry of the flame front is independent of flame quantities and scales purely with turbulence quantities.  For stable flames, the transition between propagating and material surfaces is characterized by the ratio of the  propagating velocity, $S_L^0$, and the Kolmogorov velocity scale, $u_\eta$.  This was discussed in detail by Savard and Blanquart~\cite{savard2015broken} for {\it n}-heptane/air flames, who found that reaction zones under intense turbulence fields behave as material surfaces.

\begin{figure}[tbh]
    \centering
    \includegraphics[width=1\linewidth]
    {{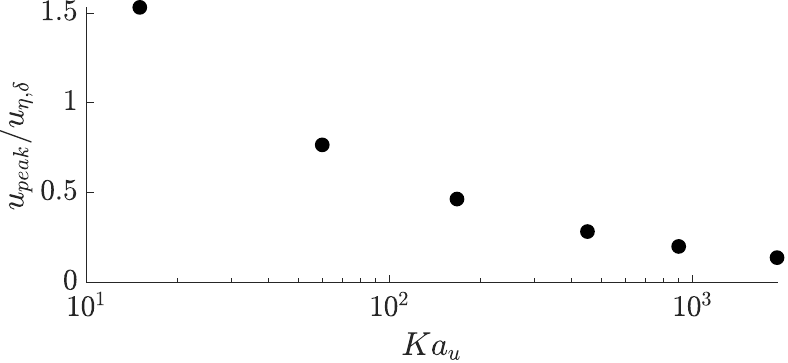}}
    \caption{Ratio of $S_L^0/u_{\eta,\delta}$ for the flames in this paper, illustrating the transition between propagating versus material surface behaviour.}
    \label{fig:material_surf}
\end{figure}

The extension to highly turbulent, unstable, lean hydrogen/air flames require more discussion.  First, there is no unique propagation speed as the thermodiffusive instabilities cause any laminar flames to have a local propagation speed varying from close to zero (local extinction) up to several times $S_L$ (for LAM case)~\cite{aspden2011characterization}.  Second, as the Karlovitz number increases, the flame may not be assumed to be infinitely thin.  For these reasons and since we are interested in the propagation of a surface located at the reaction zone, we choose $u_{peak}$, the gas velocity at the peak source term from the one-dimensional laminar flame. Likewise, the Kolmogorov velocity, $u_{\eta,\delta}$, is calculated using the viscosity at $C_{peak}$.

The ratio is shown in Fig.~\ref{fig:material_surf}. The ratio for case A is higher than one, indicating that the flame is expected to behave as a propagating surface. In this case, the intrinsic instability of the flame is expected to have a large contribution to the flame geometry. As the Karlovitz number is pushed higher, the ratio drops to be much less than one, indicating that turbulence is expected to have the dominant effect on the flame geometry.

\subsection{Flame curvature}

The curvature is calculated as the divergence of the normal vector:
\begin{equation}
    \kappa = - \nabla \cdot \mathbf{n} = \nabla \cdot \frac{\nabla C}{\vert \nabla C \vert}
\end{equation}
where the curvature is positive when the center of curvature is located in the burnt mixture. Similar to Day \textit{et al.}~\cite{day2009turbulence}, the curvature is calculated everywhere in the domain, and then sampled at an isosurface of interest, in this case, $C_{peak}$. 

\begin{figure}[tbh]
    \centering
    \subfloat[Effect of Karlovitz number]{
    \includegraphics[width=0.95\linewidth]{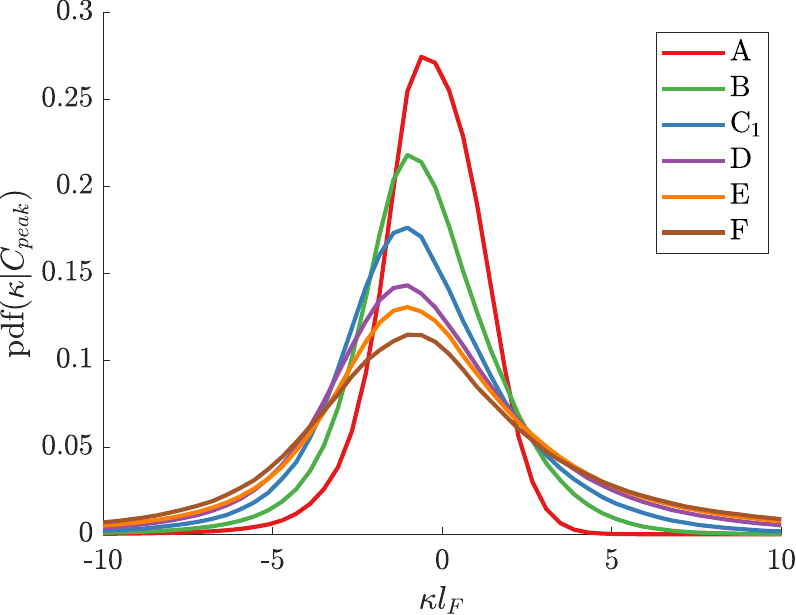}} \\
    \subfloat[Effect of integral length scale ratio]{
    \includegraphics[width=0.95\linewidth]{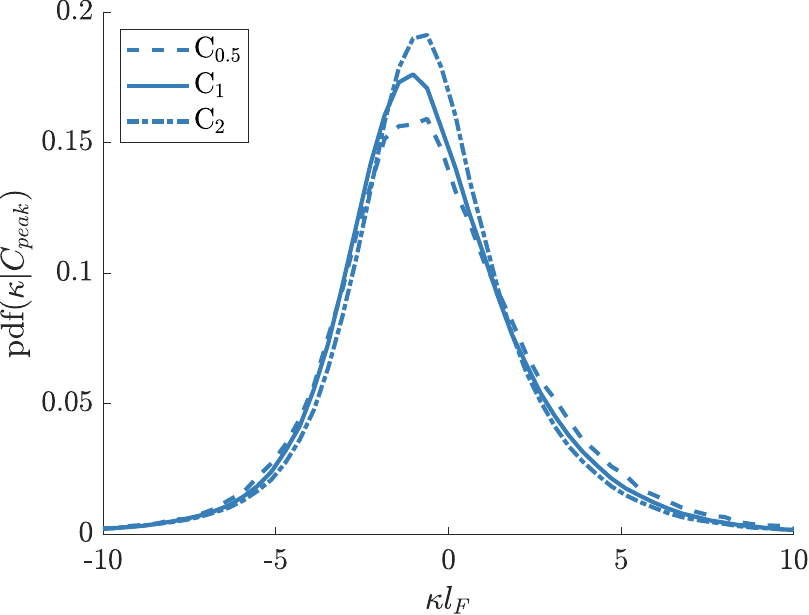}}
    \caption{Probability density function of the normalized curvature, sampled at $C_{peak}$. }
    \label{fig:pdf_curv}
\end{figure}

The probability density function of the curvature, normalized by $l_F$, and sampled at $C_{peak}$ is shown in Fig.~\ref{fig:pdf_curv}. Overall, the profiles are skewed slightly towards negative curvature, which is consistent with Lapointe and Blanquart~\cite{lapointe2016fuel}. As the Karlovitz number is increased, the peak drops by more than a factor of 2, and the variance is increased. The turbulence introduces fluctuations at smaller scales, which in turn increases the curvature. In contrast, the effect of the integral length scale (shown in Fig.~\ref{fig:pdf_curv}b) is much less severe.  The distribution narrows slightly with a higher peak value as the integral length scale is increased.  This observation reflects the negligible impact of the integral length scale/domain size on the geometry of the flame front (at a fixed Karlovitz number).

\begin{figure}[tbh]
    \centering
    \subfloat[Normalized by reaction zone thickness]{
    \includegraphics[width=0.8\linewidth]{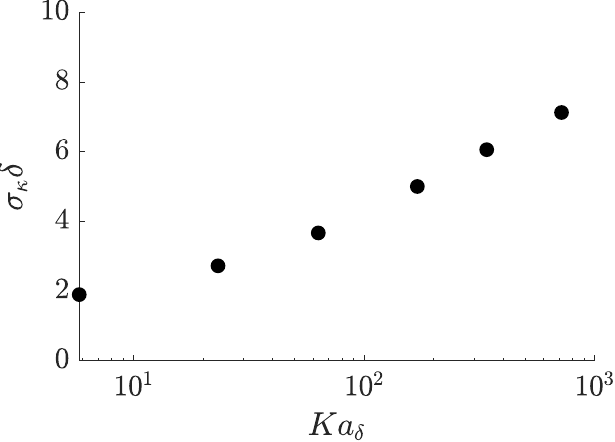}} \\
    \subfloat[Normalized by Kolmogorov length scale]{
    \includegraphics[width=0.8\linewidth]{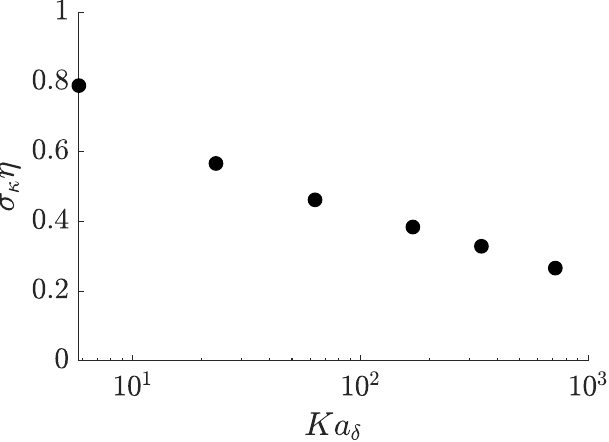}}
    \caption{Standard deviation of the curvature probability density function at $C_{peak}$, normalized by the reaction zone thickness (top) and Kolmogorov length scale (bottom).}
    \label{fig:std_curv}
\end{figure}

The spread of the curvature probability density function is measured by the standard deviation, $\sigma_\kappa$, which is shown in Fig.~\ref{fig:std_curv}. The trends are similar to those observed by Lapointe and Blanquart~\cite{lapointe2016fuel}, although the magnitudes are systematically higher, indicating relatively smaller radii of curvature compared to hydrocarbon flames at the same reaction zone Karlovitz numbers. The standard deviation appears to scale with a quantity which lies between the reaction zone thickness and the Kolmogorov length scale.

\begin{figure*}[tbh]
    \centering
    \includegraphics[width=1\linewidth]
    {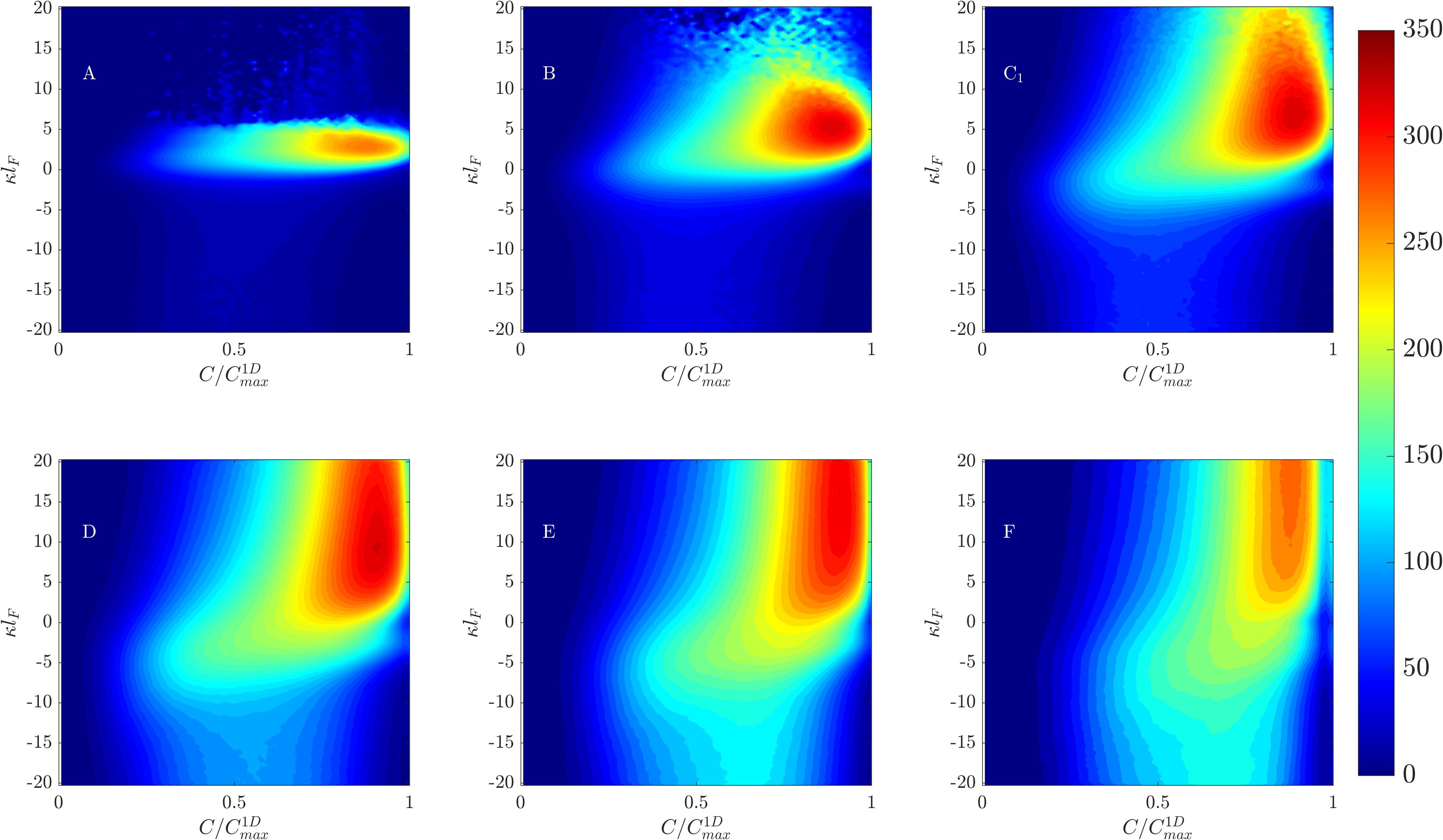}
    \caption{Two-dimensional mean source term $\langle \dot{\omega}_C \vert C, \kappa \rangle$.}
    \label{fig:2Dsrc}
\end{figure*}

\subsection{Curvature effects on source term}

\begin{figure}[tbh]
    \centering
    \includegraphics[width=1\linewidth]
    {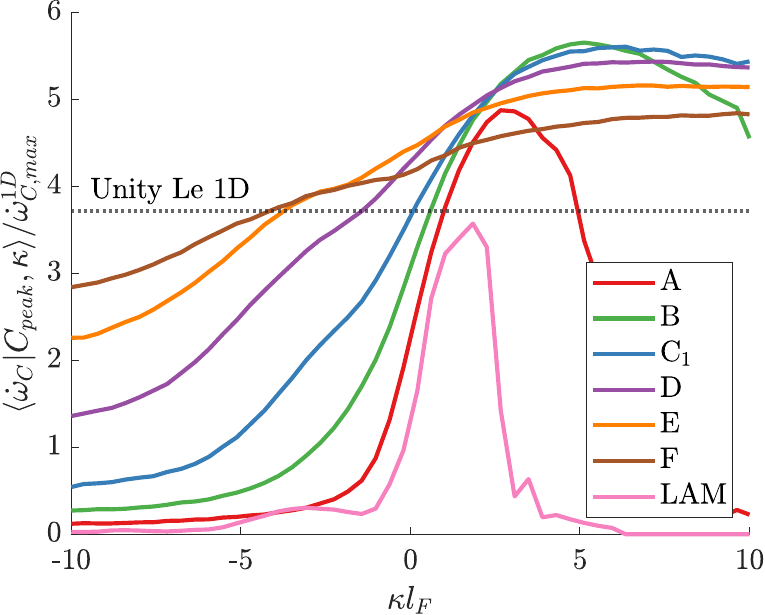}
    \caption{Mean source term conditioned on $C_{peak}$ and $\kappa$. The dotted line represents $\dot{\omega}_{C,peak}^{1D, Le}/\dot{\omega}_{C,peak}^{1D}$.} 
    \label{fig:src_curv_cpeak}
\end{figure}

The two-dimensional mean source terms conditioned on $C$ and $\kappa$ are shown in Fig.~\ref{fig:2Dsrc}. As the Karlovitz number is increased, the mean source term is spread over a larger range of $\kappa$ values. The effect of curvature on the source term appears to be dimishing with increasing $Ka_u$. To quantify this effect, 
Fig.~\ref{fig:src_curv_cpeak} shows the mean source term conditioned on $\kappa$ at $C_{peak}$. As the Karlovitz number is increased, the source term profile becomes flatter. Critically, the profiles asymptote towards the dashed line, which represents the peak source term from a one-dimensional unity Lewis flame normalized by that of the full transport flame. These figures show clearly that the effect of curvature is becoming less pronounced, which is consistent with the observation that differential diffusion effects are being dampened by the turbulence. In the limit of infinite Karlovitz number, it is expected that the effective Lewis number is unity~\cite{savard2014priori}, and the source term would become fully independent of curvature~\cite{savard2017effects}. 

\section{Conclusions}
\label{sec:conclusion}

A series of Direct Numerical Simulations of lean premixed hydrogen/air flames was conducted across a range of Karlovitz numbers and integral length scale ratios.  The results have been used to isolate the controlling parameters in the turbulent flame speed enhancement via Eq.~\eqref{eq:efficiency}. A new general expression for the burning efficiency is proposed, which is based on the conditional mean source term and gradient of the progress variable.

At a given Karlovitz number, the ratio of flame isosurface areas to the cross section area, $A(C)/A$, increases linearly with the integral length scale ratio. However, the chemical source term conditioned on the progress variable does not show significant differences. As such, the flame speed enhancement can be fully attributed to an increase in the flame area.

At a given integral length scale ratio, $\ell/l_F$, both the turbulent flame speed and area increase with the Karlovitz number before decreasing. As the Karlovitz number is increased, the shape of the mean chemical source term approaches that of the one-dimensional unity Lewis number flame, indicating that the relative importance of differential diffusion effects is lower. Consequently, the thermodiffusive instabilities are also dampened, causing a reduction in the flame area. Differential diffusion effects disappear at high Karlovitz numbers as the reaction zone is broadened and diffusivity is enhanced by the turbulence penetrating the reaction zone.

\section*{Acknowledgements}

This material is based upon work supported by the National Science Foundation under Grant No. 1832548.

This work used Stampede2 at Texas Advanced Computing Center (TACC) through allocation CTS130006 from the Advanced Cyberinfrastructure Coordination Ecosystem: Services \& Support (ACCESS) program, which is supported by National Science Foundation grants \#2138259, \#2138286, \#2138307, \#2137603, and \#2138296.

This research used resources of the National Energy Research Scientific Computing Center (NERSC), a U.S. Department of Energy Office of Science User Facility located at Lawrence Berkeley National Laboratory, operated under Contract No. DE-AC02-05CH11231 using NERSC award BES-ERCAP0023603.


\bibliographystyle{model1-num-names}

\bibliography{mybib}



\end{document}